\documentclass{article}

\pdfoutput=1
\usepackage{graphicx}
\usepackage{float}
\usepackage{amssymb}
\usepackage{amsmath}
\usepackage{bm}
\usepackage{multirow}
\usepackage[labelfont=bf]{caption}
\usepackage{mathdots}
\usepackage{bm}
\usepackage{natbib}
\usepackage{authblk}
\usepackage{epigrafica}

\title{Statistical framework for estimating GNSS bias}
\author[1]{Anthea J. Coster}
\author[1]{William C. Rideout} 
\author[1]{Philip J. Erickson}          
\author[2]{Johannes Norberg}

\affil[1]{Haystack Observatory, Massachusetts Institute of Technology,  Route 40 Westford, 01469 MA}
\affil[2]{Finnish Meteorological Institute}

\begin{document}
\maketitle

\begin{abstract}
We present a statistical framework for estimating global navigation satellite system (GNSS) non-ionospheric differential time delay bias. The biases are estimated by examining differences of measured line integrated electron densities (TEC) that are scaled to equivalent vertical integrated densities. The spatio-temporal variability, instrumentation dependent errors, and errors due to inaccurate ionospheric altitude profile assumptions are modeled as structure functions. These structure functions determine how the TEC differences are weighted in the linear least-squares minimization procedure, which is used to produce the bias estimates. A method for automatic detection and removal of outlier measurements that do not fit into a model of receiver bias is also described. The same statistical framework can be used for a single receiver station, but it also scales to a large global network of receivers. In addition to the Global Positioning System (GPS), the method is also applicable to other dual frequency GNSS systems, such as GLONASS (Globalnaya Navigazionnaya Sputnikovaya Sistema). The use of the framework is demonstrated in practice through several examples. A specific implementation of the methods presented here are used to compute GPS receiver biases for measurements in the MIT Haystack Madrigal distributed database system. Results of the new algorithm are compared with the current MIT Haystack Observatory MAPGPS bias determination algorithm. The new method is found to produce estimates of receiver bias that have reduced day-to-day variability and more consistent coincident vertical TEC values.
\end{abstract}

\section{Introduction}

A dual frequency GNSS receiver can measure the line integrated ionospheric electron density between the receiver and the GNSS satellite by observing the transionospheric propagation time difference between two different radio frequencies. Ignoring instrumental effects, this propagation delay difference is directly proportional to the line integral of electron density \citep{davies1965,vierinen2014}. 

Received GNSS signals are noisy and contain systematic instrumental effects, which result in errors when determining the relative time delay between the two frequencies. The main instrumental effects are frequency dependent delays that occur in the GNSS transmitter and receiver, arising from dispersive hardware components such as filters, amplifiers, and antennas.  Loss of satellite signal can also cause unwanted jumps in the measured relative time delay and cause unwanted non-zero mean errors in the relative time delay measurement. Because line integrated electron density is determined from this relative time delay, it is important to be able to characterize and estimate these non-ionospheric sources of relative time delay.

The non-ionospheric relative time delay due to hardware is commonly referred to as \emph{bias} in the literature. For the specific case of GPS measurements, the bias is often separated into two parts ordered by the source of delay: satellite bias and receiver bias. 

A GNSS measurement of relative propagation time delay difference including the line integrated electron density effect can be written as:
\begin{equation}
m = b + c + \int_S N_e(s) ds + \xi,
\label{eq1}
\end{equation}
where $m$ is the measurement, $b$ is the receiver bias, $c$ is the satellite bias, $S$ is the path between the receiver and the satellite, $N_e(s)$ is the ionospheric electron density at position $s$, and $\xi$ the measurement noise. The measurement is scaled to TEC units, i.e., $10^{16}/\mathrm{m}^{-2}$, and therefore bias terms also have units of TEC. See \cite{dyrud} and references therein for a further discussion. 

\begin{figure*}
\begin{center}
\includegraphics[width=0.7\textwidth]{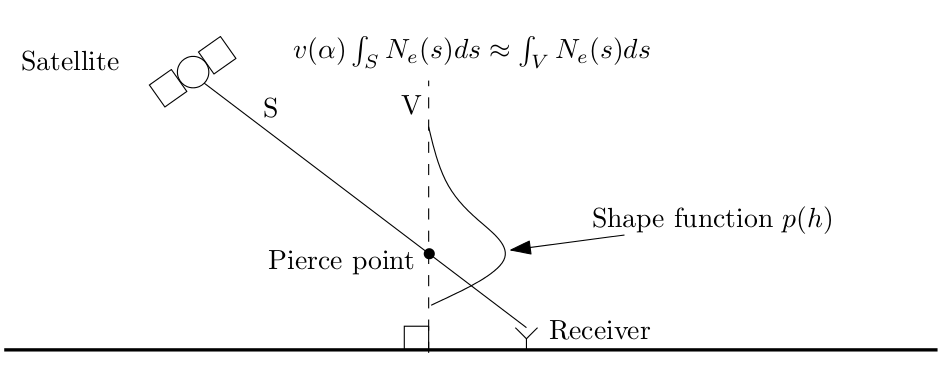}
\end{center}
\caption{A scaled altitude profile model of the ionosphere assumes that the ionosphere locally has a fixed horizontally stratified altitude profile shape multiplied by a scalar. This makes it possible to relate slanted line integrals to equivalent vertical line integrals using a elevation dependent scaling factor called the mapping function. The pierce point is located where the ray pierces the peak of the electron density profile.}
\label{slab}
\end{figure*}

For ionospheric research with GNSS receivers that perform measurements of the form shown in Equation \ref{eq1}, the quantity of interest is usually the three-dimensional electron density function $N_e(s)$. However, this quantity is challenging to derive from just GNSS measurements alone, as we only observe one-dimensional line integrals through the ionosphere. The problem is an ill-posed inverse problem called the limited angle tomography problem \citep{bust}. The difficulty arises from the fact that line integrals are measured only at a small number of selected viewing angles, and this information is not sufficient to fully determine the unknown electron density distribution without making further assumptions about the unknown measurable $N_e(s)$. These assumptions often impose horizontal and vertical smoothness, as well as temporal continuity.  

A considerable number of prior studies have attempted to solve this tomographic inversion problem in three dimensions for beacon satellites as well as for GPS satellites (see e.g. \cite{bust} and references therein). Because of the large computational costs and complexities associated with full tomographic solvers, much of the practical research is done using a reduced quantity called the vertical total electron content (VTEC).  As we will describe in more detail below, VTEC in essence results from a reduced parameterization of the ionosphere that is used to simplify the tomography problem and make it more well-posed. 
VTEC processing is only concerned with the integrated column density, and therefore the measurements are reported in TEC units. 

The fundamental assumption for vertical TEC processing is that a slanted line integral measurement of electron density can be converted into an equivalent vertical line integral measurement with a parameterized scaling factor $v(\alpha)$:
\begin{equation}
\int_{V} N_e(s) ds \approx v(\alpha) \int_{S}
N_e(s) ds,
\label{scaling}
\end{equation}
where $V$ is a vertical path, $S$ is the associated slanted path, $\alpha$ is the elevation angle, and $v(\alpha)$ is the scaling factor that relates a slanted integral to a vertical line integral. 

There are several ways that $v(\alpha)$ can be derived without resorting to full tomographic reconstruction of the altitude profile shape. Typically, the ionosphere above a certain geographic point is assumed to be described with some vertical shape profile $p(h)$ multiplied by a scalar $N_e(h) = N p(h)$. One example of an often used shape profile is the Chapman profile
\begin{equation}
p(h) = \exp\left(1-z(h)-e^{-z(h)}\right),
\end{equation}
where $z(h)=(h-h_m)/H$, $h_m$ is the peak altitude of the ionosphere, and $H$ is the scale height \citep{Feltens}. Another example is a slab with exponential top and bottom side ramps as described by \cite{coster} and \cite{Mannucci98}.  Figure \ref{slab} depicts the geometry and profile shape assumptions in vertical TEC processing.

In more advanced models, the mapping function can be parameterized not only by elevation angle but also by factors such as time of day, geographic location, solar activity, and the azimuth of the observation ray. In practice, this can be done by using a first-principles ionospheric model to derive a more physically motivated mapping function.

Although the vertical TEC assumptions described above are not as flexible as a full tomographic model that attempts to determine the altitude profile, they provide model-to-data fits that are to first order good enough to produce measurements that are useful for studies of the ionosphere. The utility of this simplified model derives from the fact that its use results in an over-determined, well-posed problem. 

The main practical difficulties in data reduction using the simplified model are estimating the receiver and satellite biases $b$ and $c$, as well as handling possible model errors. In this paper, a novel statistical framework for deriving these GNSS measurement biases is described. The method is based on examining large numbers of differences between slanted TEC measurements that are scaled with the mapping function $v(\alpha)$. The differences between pairs of measurements are assumed to be Gaussian normal random variables with a variance that is determined by the properties of the two measurements, i.e., spacing in time, geographic distance, and elevation angle. We will show how this general statistical framework can be used to estimate biases in multiple special cases and finally compare the newly presented method with an existing bias determination scheme within the MIT Haystack MAPGPS algorithm \citep{RideoutCoster}. We will refer to this new method for bias determination as Weighted linear least-squares of independent differences (WLLSID).

\section{Receiver bias estimation}

Let us denote equation \ref{eq1} in a more compact form, but now with indexing $i$ to denote the index of a measurement, $j$ to denote receiver, and $k$ to denote the satellite:
\begin{equation}
m_i = b_{j(i)} + c_{k(i)} + n_{i} + \xi_{i}.
\end{equation}
Here $n_{i}$ is the line integral of electron density through the ionosphere for measurement $i$. The receiver and satellite index associated with measurement $i$ is given by $j(i)$ and $k(i)$. Receiver noise is represented with $\xi_i$.
 
Now consider subtracting slanted TEC measurements $i$ and $i^\prime$, which are scaled with corresponding mapping function values $v_i$ and $v_{i^\prime}$, which convert slanted TEC to equivalent vertical TEC. In this analysis, it does not matter if these measurements are associated with the same receiver or the same satellite, or even if they occur at the same time. 
\begin{equation}
\begin{split}
v_{i} m_{i} - v_{i^\prime} m_{i^\prime} =  
\left(v_{i} n_{i} - v_{i^\prime} n_{i^\prime}\right) + \\
  v_{i} b_{j(i)} - v_{i^\prime} b_{j(i^\prime)} + \\
   v_{i} c_{k(i)} - v_{i^\prime} c_{k(i^\prime)} + \\
v_{i} \xi_{i} - v_{i^\prime}\xi_{i^\prime}
\end{split}
\label{diffeq}
\end{equation}

This type of a difference equation has several benefits. If measurements $i$ and $i'$ are performed at a time close to each other $t_i \approx t_{i^\prime}$ and have closely located pierce points $x_i \approx x_{i'}$, then we can make the assumption that $v_{i} n_{i} \approx v_{i^\prime}n_{i^\prime}$, i.e., that the vertical TEC is similar. 

We can statistically model this similarity by assuming that the difference of equivalent vertical line integrated electron content between two measurements is a normally distributed random variable with variance \begin{equation}
\begin{split}
v_{i} n_{i} - v_{i^\prime} n_{i^\prime} = \tilde{\xi}_{i,i^\prime} \\ \sim N\left(0,S_{i,i^\prime}\right),
\end{split}
\end{equation}
where $S_{i,i^\prime}$ is the structure function that indicates what we assume the variance of the difference of the two measurements $i$ and $i^\prime$ to be. This structure function would be our best guess of how different we expect these two measurements to be.

We assume the structure function depends on the following factors: 1) geographic distance between pierce points $d_{i,i^\prime} = |x_{i} - x_{i^\prime}|$, 2) difference in time between when the measurements were made $\tau_{i,i^\prime} = |t_i - t_{i^\prime}|$, 3) receiver noise of both measurements $\xi_{i} + \xi_{i^\prime}$, and 4) modeling errors that are dependent on elevation angles $\alpha_{i}$ and $\alpha_{i^\prime}$ of the measurements.  The modeling errors in (4) are caused by inaccuracies in the assumption that we can scale a slanted measurement into an equivalent vertical measurement. 

The following subsections describe the structure function behaviors for each dependent variable.

\subsection{Geographic distance}
In order to model the variability of electron density as a function of geographic location, we assume the difference between two measurements to be a random variable
\begin{equation}
v_i n_i - v_{i'} n_{i'} \sim N\left(0, D(d_{i,i'})\right),
\end{equation}
where in this work we use $\sqrt{D(d)} = 0.5d$ in units of $(\mathrm{TECu} / 100 \mathrm{km})$. This implies that we assume the standard deviation of difference of two vertical TEC measurements to grow at a rate of 0.5 TEC units per 100 km of spacing between pierce points.

For the results in this paper, we use the functional form above, but this can be improved in future work by a more complicated spatial structure function $D(x_i,x_{i^\prime},t_i,t_{i^\prime})$, which is a function of pierce point locations $x_i$ and $x_{i^\prime}$, as well as the time of the measurements $t_i$ and $t_{i^\prime}$. This function could for example be derived experimentally from vertical TEC measurements themselves. This would allow more accurate modeling of sunrise and sunset phenomena, as well as meridional and zonal gradients. 

\subsection{Temporal distance}

Two measurements do not necessarily have to occur at the same time, but one would expect the two measurements to differ more if they have been taken further apart from one another. This difference can also be modeled as a normal random variable
\begin{equation}
v_i n_i - v_{i'} n_{i'} \sim N\left(0, T(\tau_{i,i'})\right),
\end{equation}
where $T(\tau_{i,i'})$ is a structure function that statistically describes the difference in vertical TEC from one measurement to the other when the time difference between the two measurements is $\tau_{i,i'} = |t_i - t_{i^\prime}|$. 

In this work, we use $\sqrt{T(\tau)} = 20 \tau$ in units of $\mathrm{TECu}/\mathrm{hour}$. This makes the assumption that the standard deviation of the difference of two vertical TEC measurements grows at the rate of 20 TEC units for each hour. 

Again, an improved version of this time structure function could also be obtained by estimating it from data, but this is the subject of a future study. 

\subsection{Model and receiver errors}

There are modeling errors that are caused by our assumption that we can scale a slanted line integral to a vertical line integral as shown in Equation \ref{scaling}. First of all, this assumption does not correctly take into account that the slant path cuts through different latitudes and longitudes and thus averages vertical TEC over a geographic area. In addition to this, our mapping function assumes an altitude profile for the ionosphere that is hopefully close to reality, but never perfect. The ionosphere can have several local electron density maxima and can have horizontal structure in the form of e.g. travelling ionospheric disturbances, or typical ionospheric phenomena such as the Appleton anomaly at the equator or the ionospheric trough at high latitudes. 

In addition to this, GNSS receivers often have difficulty with low elevation measurements arising from near field multi-path propagation, which is different for both frequencies. These errors can in some cases severely affect vertical TEC estimation and thus also bias estimation.  

To first order, the errors caused by the inadequacies of the model assumptions or anomalous near-field propagation increase proportionally to the zenith angle. It is useful to include this modeling error in the equations as yet another random variable. We have done this by assuming the elevation angle dependent errors to be a random variable of the following form:
\begin{equation}
v_i n_i - v_{i^\prime} n_{i^\prime} \sim N\left(0,E(\alpha_i) + E(\alpha_{i^\prime})\right). 
\end{equation}
Here $E(\alpha_i)$ is the structure function that indicates the modeling error variance as a function of elevation angle. In this work, we use a structure function where the variance grows rapidly as the elevation angle approaches the horizon, expressed as $\sqrt{E(\alpha_i)} = 20(\cos\alpha_i)^4$. This form penalizes lower elevations more heavily. 

The structure function that takes into account vertical TEC scaling errors and receiver issues at low elevations can also be determined from vertical TEC estimates, e.g., by doing a histogram of coincident measurements of vertical TEC: 
\begin{equation}
E(\alpha) \approx \langle |\langle v_i n_i \rangle - v_{i^\prime} n_{i^\prime}|^2\rangle
\end{equation}
for all $i,i^\prime$, where $|x_i - x_{i^\prime}| < \epsilon_d$ and $|\alpha_{i^\prime}- \alpha| < \epsilon_{\alpha}$. Here $\epsilon_d$ determines the threshold for distance between pierce points that we consider to be coincidental, and $\epsilon_{\alpha}$ determines the resolution of the histogram on the $\alpha$-axis. 

\section{Generalized linear least-squares solution}

If we assume that all random variables in the structure functions of the previous section are independent random variables, we can simply add them together to obtain the full structure function
\begin{equation}
S_{i,i^\prime} = D(d_{i,i^\prime}) + T(\tau_{i,i^\prime}) + E(\alpha_i) + E(\alpha_{i^\prime}).
\end{equation}

The differences in Equation \ref{diffeq} can be expressed in matrix form as 
\begin{equation}
m = Ax + \xi,
\end{equation}
where

\begin{equation*}
 A = \begin{bmatrix}
 \ddots  & \vdots & \vdots & \vdots & \iddots \\
 \cdots  & v_{i}  & \cdots & v_{i}  & \cdots  \\
 \iddots & \vdots & \vdots & \vdots & \ddots 
 \end{bmatrix}
 -
 \begin{bmatrix}
 \ddots  & \vdots        & \vdots & \vdots &    \iddots \\
 \cdots  & v_{i^\prime}  & \cdots & v_{i^\prime}  & \cdots  \\
 \iddots & \vdots        & \vdots & \vdots        & \ddots 
 \end{bmatrix}
,
\end{equation*}
with the measurement vector containing differences between vertically scaled measurements 
\begin{equation}
m = \begin{bmatrix}
\cdots, v_{i}m_{i} - v_{i^\prime}m_{i^\prime}, \cdots
\end{bmatrix}^T
\end{equation}
and the unknown vector $x$ contains the receiver and satellite biases 
\begin{equation}
x = \begin{bmatrix}
b_0, \cdots, b_N, c_0, \cdots, c_M
\end{bmatrix}^T.
\end{equation}
For $x$, $N$ indicates the number of receivers and $M$ indicates the number of satellites. 

The random variable vector $\xi \sim N(0,\Sigma)$ has a diagonal covariance matrix defined by the structure function of each measurement pair used to form differences
\begin{equation}
\Sigma = \mathrm{diag}\left(
S_{i,i^\prime}, \cdots \right).
\end{equation}

The theory matrix $A$ forms the forward model for the measurements as a linear function of the receiver biases. 

This type of a measurement is known as a linear statistical inverse problem \citep{Jari} and it has a closed form solution for the maximum likelihood estimator for the unknown $x$, which in this case is a vector of receiver and satellite biases:
\begin{equation}
\hat{x} = (A^T \Sigma^{-1} A)^{-1} A^T \Sigma^{-1} m,
\end{equation}
This matrix equation is often not practical to compute directly due to the typically large number of rows in $A$.  However, because the matrix $A$ is very sparse, the solution can be obtained using sparse linear least squares solvers. In this work, We use the LSQR package \citep{Paige} for minimizing $|\tilde{A}x - \tilde{m}|^2$, where $\tilde{A}$ and $\tilde{m}$ are the matrix $A$ and vector $m$.  Each row of $m$ is scaled with the square root of the variance of the associated measurement $\sqrt{S_{i,i^\prime}}$ in order to whiten the noise.  In practice, this performs a linear transformation with matrix $P$ that projects the the linear system into a space where the covariance matrix is an identity matrix $P^T \Sigma P = I$.

\subsection{Outlier removal and bad receiver detection}

When a maximum likelihood solution has been obtained, a useful diagnostic examines the residuals $r = |\tilde{A}\hat{x} - \tilde{m}|$. If the residuals are larger than a certain threshold, they can be determined to be measurements that do not consistently fit the model, i.e., outliers. 

Outliers can be caused by several different mechanisms. They can be of ionospheric origin, where vertical TEC gradients are sharper than our structure function expects them to be. They can also be simply caused by a loss of lock in the receiver, which can result in a large erroneous jump in slant TEC. 

These outlying measurements can be detected and removed by a statistical test, for example $|\tilde{A}\hat{x}-\tilde{m}| > 4\sigma$, where $\sigma$ is the standard  deviation of the residuals estimated with $\sigma = \mathrm{median}\left(|\tilde{A}\hat{x} - \tilde{m}|\right)$. After the removal of problematic measurements, another improved maximum likelihood solution, one not contaminated by outliers, can be obtained. The procedure for outlier removal can be repeated over several iterations to ensure that no problematic data is used for bias estimation.


\section{Special cases}

The previous section described the general method for estimating bias by using differences of slanted TEC measurements scaled by the mapping function. However, in practice this general form rarely needs to be used. In the following sections we describe several important and practical special cases, including: known satellite bias, single receiver bias estimation, and multiple biases for each receiver.

\subsection{Known satellite bias}

If satellite bias is known a priori to a good accuracy, then it can be subtracted from the measurements and the difference equation. This reduces Equation \ref{diffeq} to
\begin{equation}
\begin{split}
v_{i} m_{i} - v_{i^\prime} m_{i^\prime} = 
\left(v_{i} n_{i} - v_{i^\prime} n_{i^\prime}\right) + \\
  v_{i} b_{j(i)} - v_{i^\prime} b_{j(i^\prime)} + 
v_{i} \xi_{i} - v_{i^\prime}\xi_{i^\prime}.
\end{split}
\label{diffeq2}
\end{equation}
This form results in the same linear measurement equations, except that the satellite biases are not unknown parameters. In this case, the theory matrix will only have at most two non-zero elements for each row. 

For GPS receivers, satellite biases are known to a good accuracy using a separate and comprehensive analysis technique \citep{Komjathy05}, and therefore this special case is approriate for bias determination for GPS receivers. 

\subsection{Single receiver and known satellite bias}

For the case that the satellite bias is known a priori and there is furthermore only one receiver, then the matrix only has one column with the unknown bias for the receiver.  

This still results in an overdetermined problem that can be solved. The solution of this special case mathematically resembles a known analysis procedure that is often referred to as "scalloping" (P. Doherty, personal communication; \cite{carrano}). This latter technique depends on the assumption that the concave or convex shape of all zenith TEC estimates collected by a single receiver observed over a 24 h period should be minimized. This same goal is obtained when time differences are minimized. The main difference in this work is that the statistical framework uses a structure function that weights differences of measurements based on time between the measurements, the elevation angle, and the pierce point distance. 

\begin{figure*}
\begin{center}
\includegraphics[width=\textwidth]{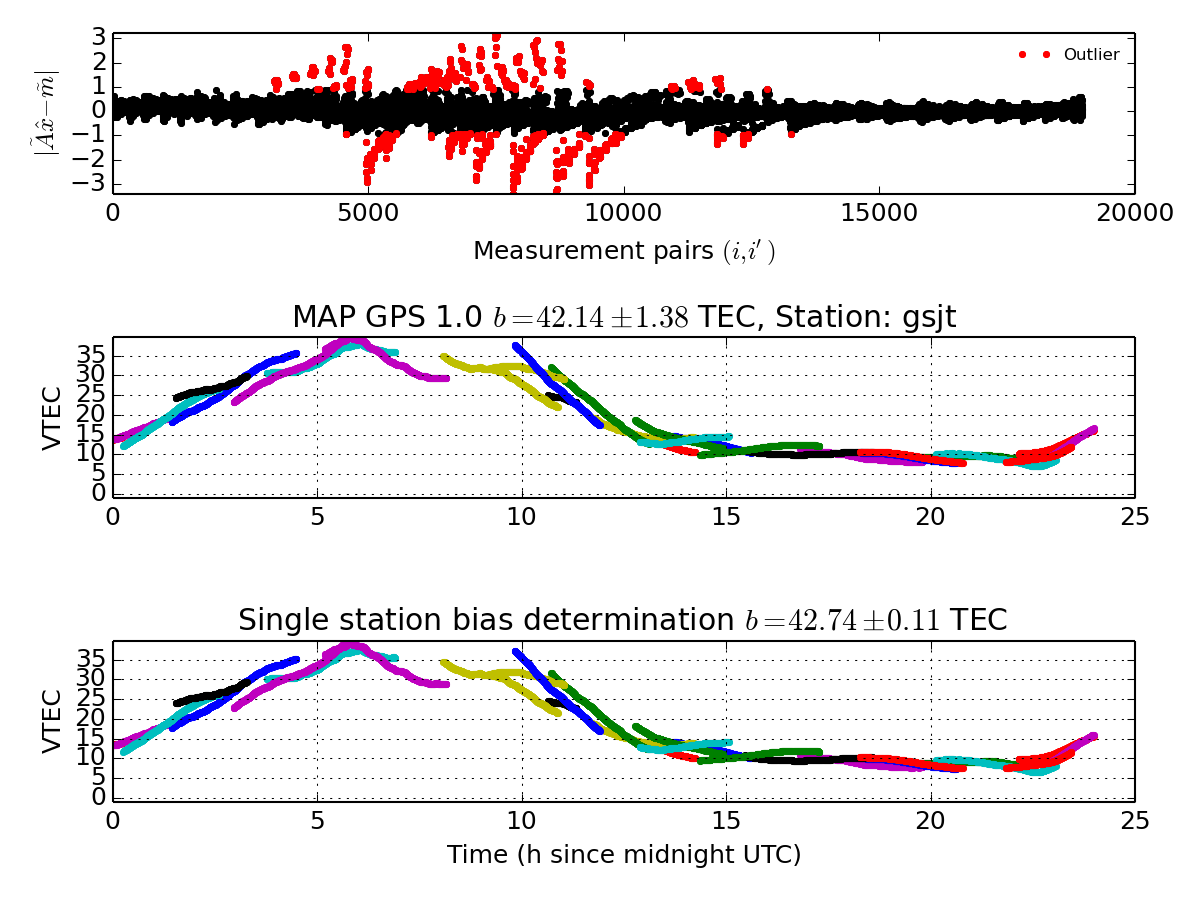}
\end{center}
\caption{Bias estimation using time differences of measurements obtained with a single receiver. Top panel shows the residuals of the maximum likelihood fit to the data. The points shown with red are automatically determined as outliers and not used for determining the receiver bias. These mostly occur during daytime at low elevations. The center panel shows vertical TEC estimated with the original MAPGPS receiver bias determination algorithm , while the bottom panel shows vertical TEC measurements obtained using only time differences using the new method described in this paper, assuming constant receiver bias and known satellite bias. The VTEC results do not differ significantly.}
\label{singlebiasex}
\end{figure*}

Figure \ref{singlebiasex} shows an example receiver bias that is determined using only data from a single receiver. In this case, time differences with $\tau_{i,i^\prime}$ less than two hours were used, in order to keep the number of measurements manageable. We also used differences of measurements between different satellites. A comparison of results with measurements obtained with the standard MAPGPS algorithm shows quite similar results between the two techniques. 

\subsection{Multiple biases}

There are several reasons for considering the use of multiple biases for the same satellite and receiver. This special case can also be handled by the same framework.

If there is a loss of phase lock on a receiver, this might result in a discontinuity in the relative time of flight measurement, which appears as a discrete jump in the slant TEC curve. Rather than attempting to realign the the curve by assuming continuity, it is possible using our framework, to simply assign an independent bias parameter to each continuous part of a TEC curve. As long as there are enough overlapping measurements, the biases can be estimated. 

For GNSS implementations other than GPS, it is possible that satellite biases are not known or cannot be treated as a single satellite bias. For example, the GLONASS network uses a different frequency for each satellite, which means that any relative time delays between frequencies caused by the receiver or transmitter hardware will most likely be different for each satellite-receiver pair. Because of this, it is natural to combine the satellite bias and receiver bias into a combined bias, which is unique for each satellite-receiver combination. 

Receiver biases are also known to depend on temperature \citep{coster2013}, because dispersive properties of the different parts of the receiver can change as a function of temperature. If an independent bias term is assigned to e.g. each satellite pass, this also allows temperature dependent effects to be accounted for, as a single satellite pass lasts only part of the day.  

Multiple bias terms can be added in a straightforward manner to the model using equation \ref{diffeq2}. This is the same equation that is used for the known satellite bias special case. Here, $b_{j(i)}$ can be interpreted as an unknown relative bias term that can vary from one continuous slant TEC curve to another. The meaning of $j(i)$ in this case is different. It is a function that assigns bias terms to measurements $i$. Each receiver doesn't necessarily need to have one unknown bias parameter, it can have many. 

\begin{figure*}
\begin{center}
\includegraphics[width=0.7\textwidth]{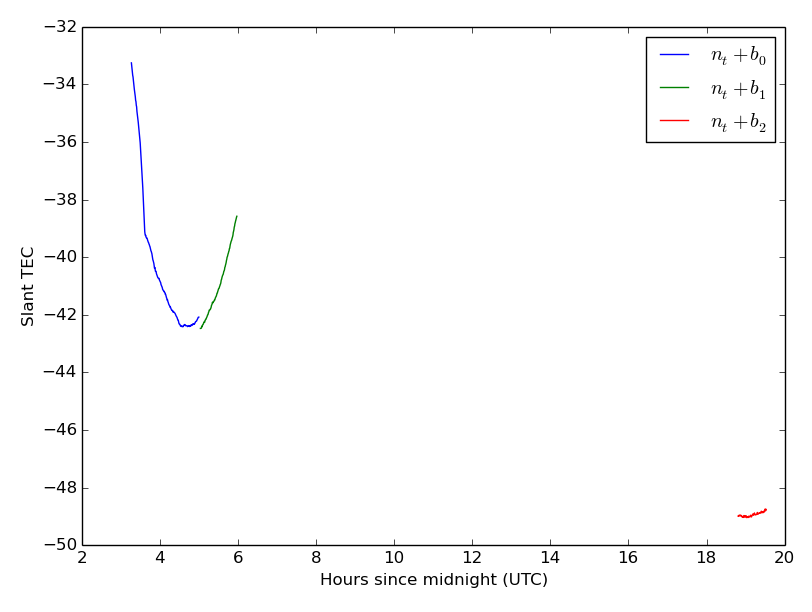}
\end{center}
\caption{An example of a measurement of a single satellite collected by a single receiver. A loss of phase lock occurs during the first pass of the satellite, resulting in two receiver biases for that pass ($b_0$, blue curve; $b_1$, green curve). During the next pass, a drift in the receiver bias could have occurred, so another receiver bias ($b_2$; red curve) is determined when the satellite is measured during the end of the day.}
\label{exfig_dis}
\end{figure*}

An example of a measurement where the same satellite is observed using a single receiver is shown in Figure \ref{exfig_dis}. In this case, the satellite is measured in the morning first, and during the pass, there is a discontinuity in the TEC curve, most likely due to loss of lock. We give the measurements before $\sim$5 UTC and from $\sim$5 to 6 UTC an independent bias term $b_0$ and $b_1$. The same satellite is seen again in the evening at 19 UTC, and we again assign a new bias term to it $b_2$. 

\begin{figure*}
\begin{center}
\includegraphics[width=0.8\textwidth]{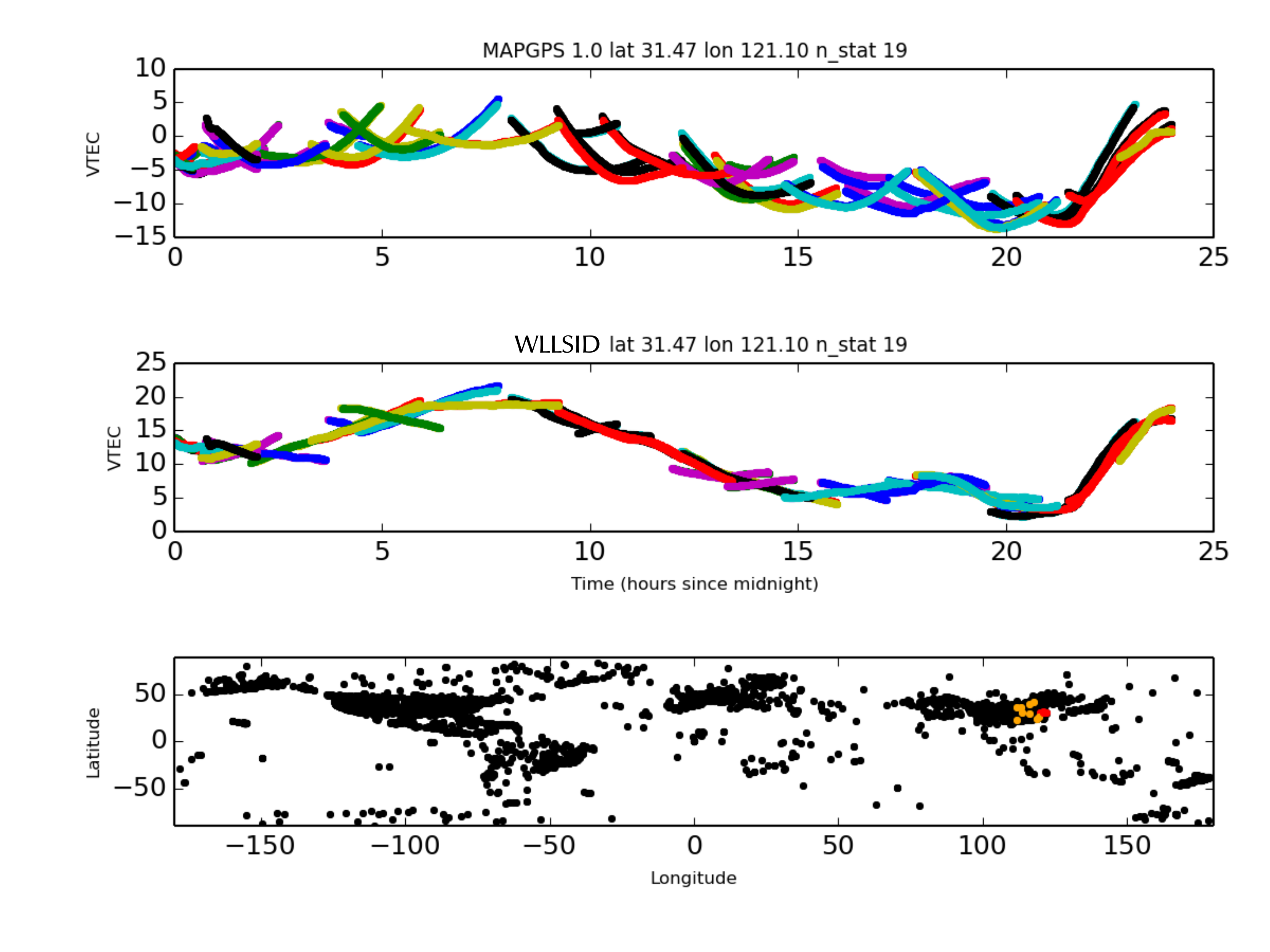}
\end{center}
\caption{Vertical TEC with satellite bias estimated using the current version of the MIT Haystack Observatory MAPGPS algorithm (Rideout and Coster 2006) shown above. Multiple receivers have problems with receiver stability, which makes the assumption of unchanging receiver bias problematic and causes the receiver bias determination to fail. Vertical TEC with receiver biases obtained using the multiple biases assumption is shown below. The new method produces a more consistent baseline. The red dots show stations that are plotted. The algorithm uses all of the data from the 19 stations marked with orange and red dots. The stations marked with orange are used to assist in reconstruction by using a larger geographic area.}
\label{exfig_china}
\end{figure*}

Another multiple bias example is shown in Figure~\ref{exfig_china}, which displays measurements from 19 neighboring receivers in China. A few of these receivers have discrete jumps in the slanted TEC curves that make it impossible to assume a constant receiver bias during the course of the entire day. This can be seen as a poor fit using the standard MIT Haystack MAPGPS algorithm. When multiple bias terms are introduced (in the same way as depicted in Figure \ref{exfig_dis}), the measurements from these stations can be recovered.

\section{Comparison}


In order to test the framework in practice for a large network of GPS receivers, we implemented the framework described in this paper as a new bias determination algorithm for the MIT Haystack MAPGPS software, which analyses data from over 5000 receivers on a daily basis. We used the MAPGPS program to obtain slant TEC estimates. Then, instead of using the MAPGPS routines for determining receiver biases, we used the new methods described in this paper. We label results obtained using the new bias determination algorithm with WLLSID. 

When fitting for receiver bias, we assumed a fixed receiver bias for each station over 24 hours. We also assumed a known satellite bias, which was removed from the slant measurement. To keep the size of the matrix manageable, we selected sets of 11 neighboring receiver stations and considered each combination of measurements across receiver and satellites occurring within five minutes of each other as differences that went into the linear least squares solution. For this comparison, we did not use time differences.

To estimate the goodness of the new receiver bias determination, we compared the method with the existing MAPGPS algorithm for determining receiver bias, which utilizes a combination of scalloping and differential linear least-squares \citep{RideoutCoster,gaposchkin}.

\subsection{Self consistency comparison}

As a measurement of goodness, we used the absolute difference between two simultaneous geographically coincident measurements of vertical TEC $|v_i n_i - v_{i^\prime}n_{i^\prime}|$. The two measurements were considered coincident if the distance between the pierce points was less than 50 km and the measurements occurred within 30 seconds of each other. We also required that the two measurements were not obtained using the same receiver. As a figure of merit, we used the mean value of the absolute differences:
\begin{equation}
F=\frac{1}{N} \sum_{i\ne i^\prime} |v_i n_i - v_{i^\prime}n_{i^\prime}|
\end{equation}
This figure of merit measures the self-consistency of the measurements, i.e., how well do the vertical TEC measurements obtained with different receivers agree with one another. The smaller the value, the more consistent the vertical TEC measurements are. 

All in all, we found 192360 such coincidences for the 5220 GPS receivers in the database on 24 hour period starting at midnight 2015-03-17. Biases for the measurements were obtained both with the new and existing MAPGPS bias determination methods (MAPGPS and WLLSID). The figure of merit for the existing MAPGPS method was 2.25 TEC units, and the WLLSID method has a figure of merit of 1.62 TEC units, which is about 30 \% better. 

\begin{figure*}[h!]
\begin{center}
\includegraphics[width=\textwidth]{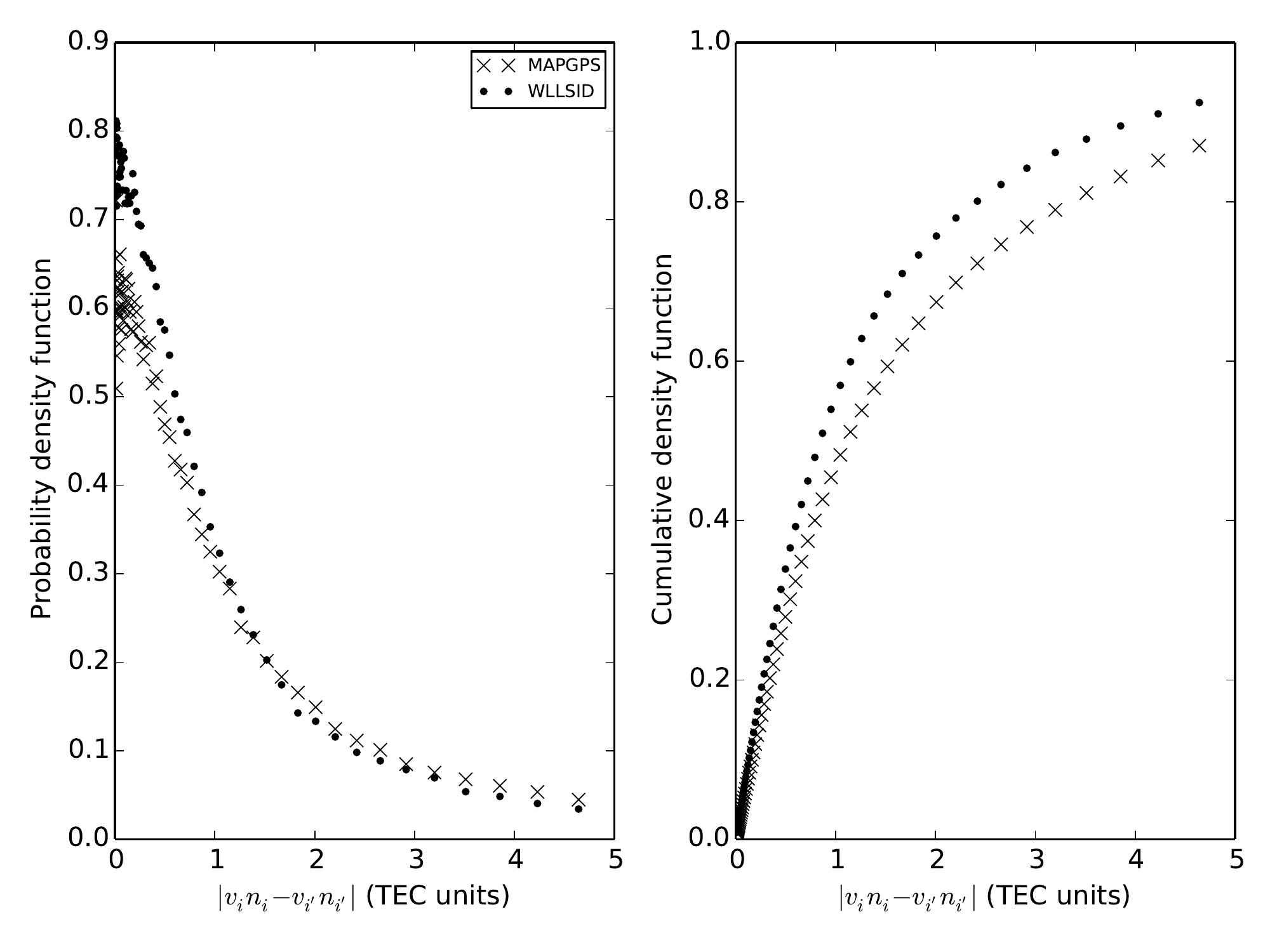}
\end{center}
\caption{Probability density function and cumulative density functions for 192360 coincidences where vertical TEC was measurement within the same 30 second time interval and have pierce points less than 50 km apart from one another. The new method (labeled as WLLSID) has significantly less $< 1$ TEC unit differences}
\label{compfig2}
\end{figure*}

The probability density function and cumulative density function estimates for the coincident vertical TEC differences are shown in Figure \ref{compfig2}. The new method results in significantly more $< 1$ TEC unit differences than the old method. It is evident from the cumulative distribution function that both methods also result in some coincidences that are in large disagreement with each other. The result occurs at least in part due to our inclusion of elevations down to 10 degrees in the comparison, and it is therefore expected that some low elevation measurements will be significantly different from one another. 

\subsection{Receiver bias day-to-day change}

We also investigated receiver bias variation from day to day. We arbitrarily selected two consecutive quiet days: Days 140 and 141 of 2015. We calculated the sample mean day-to-day change in receiver bias across all receivers:
\begin{equation}
\delta b = \frac{1}{N} \sum_{i=0}^{N} b_{i,140} - b_{i,141},
\end{equation}
where $N$ is the number of receiver. In addition to this, we calculated the standard deviation using sample variance:
\begin{equation}
\sigma_b = \sqrt{\frac{1}{N-1} \sum_{i=0}^{N} \left( (b_{i,140} - b_{i,141})-\delta b\right)^2}.
\end{equation}
For the MAPGPS method, we found overall that $\delta b = -0.2 \pm 0.05 (2\sigma)$ TEC units and $\sigma_b = 1.6$ TEC units. With the new WLLSID method, we found that $\delta b = 0.02 \pm 0.05 (2\sigma)$ TEC units and $\sigma_b = 1.3$ TEC units. This indicates that not only is the day to day variability slightly smaller with the new method, but also that the old method has a statistically significant non-zero mean day-to-day change in receiver bias, which is not seen with the new method. When the data is broken down into high and equatorial latitudes, the result is  similar.



\section{Conclusions}

In this paper, we describe a statistical framework for estimating bias of GNSS receivers by examining differences between measurements. We show that the framework results in a linear model, which can be solved using linear least squares. We describe a way that the method can be efficiently implemented using a sparse matrix solver with very low memory footprint, which is necessary when estimating receiver biases for extremely large networks of GNSS receivers. 

We compare our method for bias determination with the existing MIT Haystack MAPGPS method and find the new method results in smaller day-to-day variability in receiver bias, as well as a more self-consistent vertical TEC map. 

The weighting of the measurement differences is done using a structure function. We outline a few ways to do this, but these are not guaranteed to be the best ones. Future improvements to the method can be obtained by coming up with a better structure function, which can possibly be determined from the data itself, e.g., using histograms, empirical orthogonal function analysis, or similar methods.

While we describe how differences result in a linear model, we do not explore to a large extent in this work the possible ways in which differences can be formed between measurements. Because of the large number of measurements, obviously all the possible differences cannot be included in the model. In this study, we only explored two types of differences: 1) differences between geographically separated temporally simultaneous measurements obtained with tens of receivers located near each other, 2) differences in time less than two hours performed with a single receiver. There are countless other possibilities, and it is a topic of future work to explore what differences to include to obtain better results. 

We describe several important special cases of the method: known satellite bias, single receiver and known satellite bias, and the case of multiple bias terms per receiver. The first two are applicable for GPS receivers and the last one is applicable to GLONASS measurements, as well as measurements where a loss of lock as caused a non-zero mean step-like error in the TEC curve. 

\section{Acknowledgments}

GPS TEC analysis and the Madrigal distributed database system are supported at MIT Haystack Observatory by the activities of the Atmospheric Sciences Group, including National Science Foundation grants AGS-1242204 and AGS-1025467 to the Massachusetts Institute of Technology.  Vertical TEC measurements using the standard MAPGPS algorithm are provided free of charge to the scientific community through the Madrigal system at {\tt http://madrigal.haystack.mit.edu}.

\bibliographystyle{natbib}

\end{document}